\documentclass[twocolumn]{aastex62}
\usepackage{subfigure}
\usepackage[utf8]{inputenc}
\usepackage{graphicx}
\usepackage{dcolumn}
\newcommand{\NumGW}{11 }
\begin{document}
\title{Where to find Electromagnetic Wave Counterparts of stellar-mass binary black hole mergers?}
\author{Shu-Xu Yi}
\altaffiliation{Department of Astrophysics, Radboud University Nijmegen, \\
P.O. Box 9010, NL-6500 GL Nijmegen, The Netherlands\\}

\author{K.S. Cheng}
\altaffiliation{Pokfulam Road, Department of Physics, \\
the University of Hong Kong, Hong Kong, China\\}



\begin{abstract}
Multi-messenger astronomy combining Gravitational Wave (GW) and Electromagnetic Wave (EM) observation brings huge impact on physics, astrophysics and cosmology. However, the majority of sources to be detected with currently running ground-based GW observatories are binary black hole (BBH) mergers, which are expected disappointedly to have no EM counterparts. In this letter, we propose that if the BBH merger happens in a gaseous disk around a supermassive black hole, the merger can be accompanied by a transient radio flare alike a fast radio burst (FRB). We argue that the total mass and the effective spin derived from GW detection can be used to distinguish such a source from other channels of BBH mergers. If the prediction is confirmed with future observation, multi-messenger astronomy can be brought to a distance which is one order of magnitude farther than present. The mystery of the origin of FRBs can also be revealed partially.

\end{abstract}


\section{Introduction}\label{sec:intro}
Since the first Gravitational Wave (GW) event detected in 2015, there have been \NumGW published events (LIGO/Virgo science runs O1-O2, \citealt{2018arXiv181112907T}), and this number is increasing at an accelerated rate in the current O3 run (https://gracedb.ligo.org/latest/). The observation of the Electromagnetic Wave (EM) counterpart of GW170817 revealed a promising prospect of multi-messenger in fundamental physics (e.g. \citealt{2017PhRvL.119y1301B,2018PhRvD..98l4018N}), astrophysics (e.g. \citealt{2018PhRvL.121p1101A,2019PhRvL.122f2701W}) and cosmology \citep{2018PhRvL.121b1303V}. Among targets of LIGO/Virgo and other ground-based GW detectors, double neutron stars (DNS) mergers, and more rarely neutron star-black hole (NBH) mergers are only kinds of sources which are expected to be accompanied with EM counterparts. It means that the majority of targets, i.e., mergers of stellar-mass binaries black holes (BBH) were thought hopeless to be detected via EM. As a result, the follow-up searching for the EM counterpart is mainly focused on a GW detection which shows high possibility to be a DNS or a NBH. On the other hand, if some of the BBH merger can also be detected with EM, 
multi-messenger astronomy can be studied at a distance which is one order of magnitude farther than DNSs. The sample will also be increased significantly. 

There are three well-known channels of forming coalescing BBHs: 1, isolated evolution of massive binaries via common envelope phase \citep{2013A&ARv..21...59I,2014LRR....17....3P}; 2, isolated evolution of massive binaries via rotational chemical mixing \citep{2016MNRAS.460.3545D,2016MNRAS.458.2634M,2016A&A...588A..50M}; 3, dynamical capture in dense clusters (see \citealt{2017ApJ...834...68C} and references therein). The above mentioned channels produce BBH mergers in ``clean" environments, where no electromagnetic radiation is expected \citep{2016arXiv160402537Z}. We refer to those channels as ``clean channels" (See \cite{2018arXiv180605820M} for a pleasant review).  
\begin{figure}[ht]
    \centering
    \includegraphics[trim={3cm 0 7.5cm 0},clip,width=0.45\textwidth]{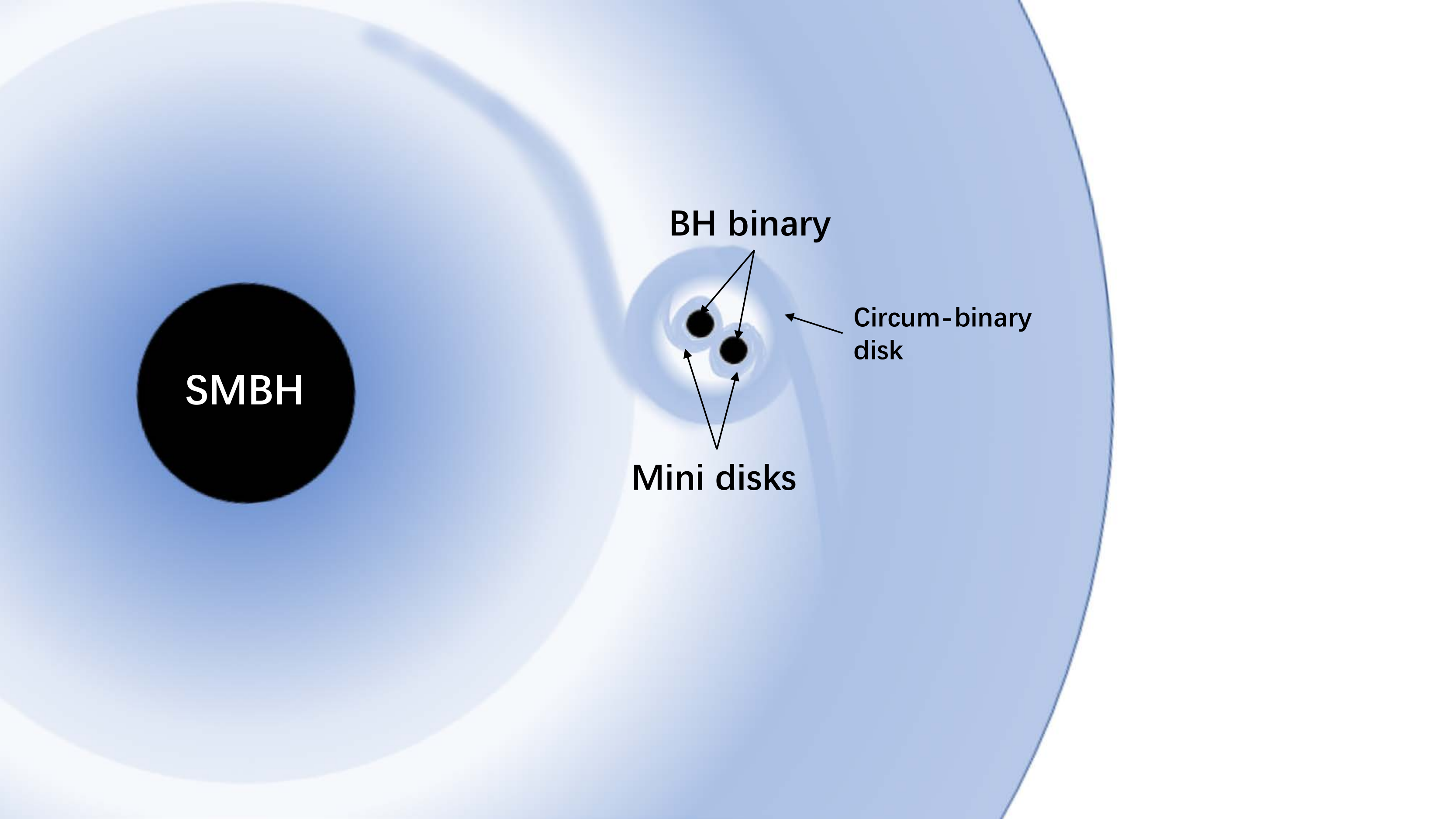}
    \caption{Illustration of a binary black hole system embedded in a accretion disk surrounding a supermassive black hole.}
    \label{fig:illustration}
\end{figure}
Recently, a new channel of producing BBH mergers has received attention, i.e., BBHs embedded in the disk surrounding a supermassive black hole (SMBH), or an active galactic nucleus (AGN). We refer to this channel as ``the AGN-disk channel" \citep{2017ApJ...835..165B,2017MNRAS.464..946S,2018MNRAS.474.5672L,2018ApJ...859L..25Y}. In this circumstance, some authors predicted that there would be EM radiation accompanying GW. \cite{2010PhRvD..81h4008F} found that the shock due to the orbital motion of the binary could give X-ray radiation. Such a radiation may not be luminous enough to be detected from a distance larger than several hundreds Mpc. \cite{2017ApJ...835..165B} suggested there is thermal emission from the transient accretion disk of the BBH and/or Doppler boosted emissions from the relativistic outflow. The fluxes of those high energy (HE) emissions would be well under the limit of current detectors, unless the accretion rate is highly super-Eddington. 

In difference with the previous studies that focused on HE radiation, in this letter we propose that these BBH mergers will give flares in radio bands (several hundreds MHz to GHz), alike those in Fast radio bursts (FRB) but with a bit longer duration. The spectral flux density of such radio flares are above the detecting limit of radio telescopes with large collecting area. In section \ref{sec:frb}, we will present the emission mechanism and give predictions on the EM counterparts. 

In the mean time, it is beneficial to ask the following question: how to distinguish between the AGN-disk channel BBH-originating GW events and ones from clean channels, solely with GW observation? Here we suggest that, their total masses and effective spin parameters ($M_{\rm{tot}}$ and $\chi_{\rm{eff}}$) can serve as indicators of their origin. The simulation of \cite{2019MNRAS.483.3288P} shows that BBH mergers from isolated massive binary evolution have their $\chi_{\rm{eff}}$ clustering around zero. For the dynamical channel, similar distribution is expected. In the meanwhile, the AGN-disk channel BBH would have opportunity to accrete mass and angular momenta from surrounding materials. As a result, we expect them to possess larger $M_{\rm{tot}}$ and $\chi_{\rm{eff}}$. Indeed, GW170729 has the largest $M_{\rm{tot}}$, whose $\chi_{\rm{eff}}$ also stands out of the cluster around zero where others lie. Hence, we suspect that GW170729 could be the first example of the AGN-disk channel BBH mergers. 

In section \ref{sec:parameterspace}, we give the probability density distribution of $M_{\rm{tot}}$ and $\chi_{\rm{eff}}$ of BBH mergers which are originated from the AGN-disk channel. Analytic formalism and Monte Carlo simulation are used in this study. In section \ref{sec:discussion}, we discuss the observational aspect in the searching for EM counterpart of GW from BBH and their host galaxies. We also discuss caveats of our results. We use geometric unit ($G=c=1$), unless otherwise indicated. 

\section{Transient radio flares as EM counterparts of BBH mergers from the AGN-disk channel}\label{sec:frb}
The scenario that we are studying in this letter is the same as that in previous literature: a BBH, either formed in or migrated to and trapped in the disk around a SMBH, inspirals and coalesces under the assistance of accretion from the surrounding matter. As shown in previous numerical simulations, several features are expected (see figure \ref{fig:illustration} for an illustration): a region with less dense gas in the AGN-disk is cleaned by the BBH. It can be a gap-ring in the orbit where the BBH lies, or a cavity around the BBH \citep{2011ApJ...726...28B}. Gas inflows through over-dense streams from the AGN-disk, forming a circum-binary disk (CD). Some of the gas in the CD are transferred onto individual BHs via their mini-disks. While other gas in the CD are barricaded from being accreted due to the slingshot effect of the binary orbit, which gives rise to an inner gap in the CD. 

After the coalescence of the BBH, remnant materials in mini-disks and those accumulated in the CD will be quickly accreted onto the newly formed, highly spinning Kerr BH. The accretion could be transient and temporarily super-Eddington, analogy to that in tidal disruption events \citep{2013ApJ...767...25G,2016Natur.535..388K,2018MNRAS.478.3016W} or micro-quasars \citep{2014Sci...343.1330S}. As proposed in \cite{2019MNRAS.483.4197Y}, a transient high mass transfer onto a stellar mass BH from its stellar companion could trigger a clumpy jet, from where fast radio bursts (FRB) could arise. Similar scenario also applies here: The transient accretion of the remnant gas onto the newly formed Kerr BH can launch an inhomogeneous Blandford-Znajek jet. Gaseous clumpy ejecta in the jet have diverse velocities and are ejected at different instances. Therefore, they might collide among each other at some distance $l_{\rm{col}}$. $l_{\rm{col}}$ is related to the time interval between the ejected instances $\delta t$ and the average Lorentz factor of the plasma bulk motion $\gamma$:
\begin{equation}
    l_{\rm{col}}\sim c\gamma^2\delta t.
\end{equation}
We denote the height of the transient accretion disk at the inner most radius as $h$. The typical separation between clumps being accreted is in the same order of magnitude of $h$. Therefore $\delta t\sim h/v$, where $v$ is the free-falling velocity at the inner most radius. 

The collision between clumps in the jet could trigger plasma oscillation, which will form temporarily charge-separated bunches. The plasma frequency of the oscillation is:
\begin{equation}
   \nu_{\rm{pls}}(l)=2\sqrt{\frac{\gamma e^2n_{\rm{e}}(l)}{\pi m_{\rm{e}}}},\label{eqn:plasmaFre}
\end{equation}
where $n_{\rm{e}}(l)$ is the number density of electrons, which is simply related to the mass density of the fully ionized plasma $\rho(l)$. Since the dimension of the clump expands along the jet with increasing cone radius, the density of the plasma decreases accordingly as:
\begin{equation}
    \rho(l)=\rho_0\left(\frac{h}{s(l)}\right)^3,
\end{equation}
where $\rho_0$ is the mass density of the gas at the inner most radius, which can be related to the accretion rate with:
\begin{equation}
    \dot{M}=2\pi r_{\rm{in}}h\rho_0v.
\end{equation}

Taking above equations into equation (\ref{eqn:plasmaFre}), we obtain the plasma frequency of the ejecta as function of the distance to the BH:
\begin{equation}
\nu_{\rm{pls}}(l)=3.45\sqrt{\frac{f_{\rm{Edd}}}{m_\bullet\eta\tilde{h}}}\gamma^{-2.5}_{100}\left(\frac{l_{\rm{col}}}{\Theta_{0.1}l}\right)^{1.5}\,\text{GHz}.
\end{equation}
In the above equation, $f_{\rm{Edd}}$ and $\eta$ are the Eddington ratio and the radiation efficiency of accretion respectively. 

When such a bunch slides along the curling magnetic field lines in the jet, curvature radiation will be emitted. The frequency of the curvature radiation in the rest frame is:
\begin{equation}
    \nu_{\rm{cur}}(l)=2\gamma_{\rm{los}}\gamma^3_\parallel c/s(l),
\end{equation}
where $\gamma_{\rm{los}}$ is the Lorentz factor of the bulk motion along the line-of-sight, $\gamma_\parallel$ is the Lorentz factor corresponding to the sliding of the plasma along the magnetic field lines. For simplicity, we assume $\gamma_{\rm{los}}\sim\gamma_\parallel\sim\gamma$; $s(l)$ is the local radius of curvature of the magnetic field lines. Since the magnetic field lines are highly spiral in the jet, we can relate $s(l)$ approximately with the opening angle of the cone of jet as: $s(l)\sim\Theta l$. 

As a result, the curvature radiation frequency as function of distance from the BH is: 
\begin{equation}
\nu_{\rm{cur}}(l)=3.8\tilde{h}^{-1}\frac{\gamma_{100}^2}{m_\bullet}\frac{l_{\rm{col}}}{\Theta_{0.1}l}\,\text{GHz},
\label{eqn:nucur}
\end{equation}
where $\tilde{h}$ is the height of the transient accretion disk scaled with the inner most radius $r_{\rm{in}}$, $\gamma_{100}$ is the Lorentz factor of the bulk motion of the bunch scaled with 100, $m_\bullet$ is the mass of the merged BH in unit of $M_\odot$, $\Theta_{0.1}$ is the opening angle of the jet cone divided by 0.1. 

When the condition that $\nu_{\rm{cur}}=\nu_{\rm{pls}}$ meets at some distance where $l>l_{\rm{col}}$, the plasma instability will grow, and the kinetic energy of the bunch will be coherently emitted in a $\mu$s time scale. It corresponds to a spike in the radio light curve. Each collision among the jet clumps has the opportunity to give such a spike, and thus we expect a multiple components radio flare, which is composed by many spikes, alike those sub-structures seen in FRBs. The clumpy jet is thought to arise due to the sudden increase of accretion rate at the moment of merger. The follow-up accretion is expected to be less temporal varying, and the accretion rate is as high as in the onset.

The duration and the power of the flare are proportional to the mass of the BH. The former corresponds to the free-falling time scale near the inner most stable orbit (ISCO): 
\begin{equation}
\tau_{\rm{dur}}\sim r_{\rm{ISCO}}/c.
\end{equation}
For a BH with $60\,M_\odot$, $\tau_{\rm{dur}}\approx1.8\,$ms.  

The apparent luminosity is:
\begin{equation}
L=\frac{8\eta m_\bullet f_{\rm{Edd}}}{2\pi(1-\cos\Theta)}\times10^{38}\,\text{ergs/s}.
\label{eqn:L}
\end{equation}
With typical mass of coalescent BHs and a super-Eddington accretion rate, $L$ can be $10^{41}-10^{42}\,\text{ergs/s}$. 
Therefore, we shall expect the EM counterparts to be longer and brighter FRBs than usual. Such a FRB, if localized with high accuracy, is expected to be found at core region of an active galaxy, and be simultaneous with a GW chirp. We encourage the reader to refer to \cite{2019MNRAS.483.4197Y} for more detailed derivation and discussion on the physics of this model.

\section{The effective-spin and total mass of BBH in the AGN-disk channel} \label{sec:parameterspace}
The spin evolution of a BH under accretion along the equatorial plane is first investigated by \citep{1970Natur.226...64B}. His result did not take into account the torque exerted by photons emitted from the accretion disk. This term will cause a correction to Bardeen's equation, making the extreme value of $a^*=0.998$ instead of unity \citep{1974ApJ...191..507T}. We neglect this small correction here. When the angular momentum of the accretion disk is misaligned with the spin of the BH, the evolution of the BH angular momentum is \citep{2009MNRAS.399.2249P}:
\begin{equation}
    \frac{d\mathbf{J}_{\rm{BH}}}{dt}=\dot{M}L_{\rm{ISCO}}\hat{\mathbf{l}}+4\pi\int_{\rm{disk}}\frac{\mathbf{L}(R)\times\mathbf{J}_{\rm{BH}}}{R^2}dR, \label{eqn:spinchange}
\end{equation}
where $L_{\rm{ISCO}}$ is the specific angular momentum brought onto the black hole from ISCO; $\hat{\mathbf{l}}$ is a unit vector, which is guaranteed to parallel with $\mathbf{J}_{\rm{BH}}$ due to the Bardeen-Petterson effect \citep{1975ApJ...195L..65B}; $\mathbf{L}(R)$ is the angular momentum in per unit area in the accretion disk in the distance $R$ from the BH. The first term on the right is the change rate of spin modulus, and the second term is the change rate of spin direction. 

Denote that
\begin{equation}
    a_*=\pm\frac{J_{\rm{BH}}}{M^2},
\end{equation}
where $J_{\rm{BH}}$ is the modulus of the spin angular momentum. When the spin is counter rotating with the binary orbit, the signature of $a_*$ is minus. The evolution of $a_*$ just follow the Bardeen's equation:
\begin{equation}
    \frac{da_*}{d\ln M}=\frac{1}{M}\frac{L_{\rm{ISCO}}}{E_{\rm{ISCO}}}-2a_*,
\label{eqn:differential}
\end{equation}
where $E_{\rm{ISCO}}$ is the specific energy which was brought onto the black hole from ISCO. With the explicit expression of $L_{\rm{ISCO}}$ and $E_{\rm{ISCO}}$ \citep{1972ApJ...178..347B}, the above equation can be integrated to give $a_*$ as function of $x$ and initial $a_{*,0}$, where $x\equiv M_{\rm{f}}/M_0$ is the ratio between the mass after accretion and the initial mass of the BH. The explicit expression of $a_*(a_{*,0},x)$ is lengthy, thus we just use the numerical integration of equation (\ref{eqn:differential}) for our purpose. 


The second term on right-hand side of equation (\ref{eqn:spinchange}) governs the alignment of BH spin and angular momentum of the disk. The angle between the two vector $\theta$ decline to zero exponentially in an alignment time scale:
\begin{equation}
    \theta(t)=\theta_0\exp\left(-\frac{t}{\tau_{\rm{align}}}\right),
\end{equation}
where $\theta_0$ is the initial $\theta$ at $t=0$, and $\tau_{\rm{align}}$ is the alignment time scale, which is:
\begin{equation}
    \tau_{\rm{align}}=1.13\times10^5\alpha^{58/35}_{0.1}f^{-5/7}_{\nu_2}M^{-2/35}_6\left(\frac{f_{\rm{Edd}}}{\eta_{0.1}}\right)^{-\frac{32}{35}}a_*^{5/7}\,\rm{yr}.
\end{equation}
Since we denote the parallel spin with positive $a_*$ and anti-parallel spin with negative $a_*$, the range of $\theta$ is from 0 to 90$^\circ$. 

In the above equation, $\alpha_{0.1}$ is the $\alpha$ parameter in a standard thin disk divided by 0.1 \citep{1973A&A....24..337S}; $f_{\nu_2}$ is a non-linear-effect-related coefficient defined from simulation; $M_6$ is the mass of BH in unit of $10^6\,M_\odot$ and $\eta_{0.1}\equiv\eta/0.1$. Those parameters are all in the order of unity, therefore we set them as one in the following Monte-Carlo simulation. 

The equation (9) of \cite{2018ApJ...859L..25Y} gives the mass of individual BH as function of its initial mass, orbital separation and properties of the AGN disk:
\begin{equation}
    m_{\rm{f}}=\left(\frac{a^4_{12,0}f_{\rm{Edd}}\gamma m_0^{4\gamma}}{3.94\times10^{-6}}\right)^{1/(3+4\gamma)},
    \label{eqn:2018}
\end{equation}
where $m_{\rm{f}}$ and $m_0$ are the masses of the individual BH in unit of $M_\odot$, at coalescence and in initial respectively; $a_{12,0}$ is the initial orbital separation of the BBH in unit of $10^{12}$\,cm; $f_{\rm{Edd}}$ is the Eddington ratio of accretion onto individual BH during inspiral; $\gamma$ a parameter used in the numerical simulation of \cite{2017MNRAS.469.4258T} to describe the mass sink rate. In the above equation, we fix $\eta_{0.1}=1$, thus absorb any change of $\eta$ in to the definition of $f_{\rm{Edd}}$. In \cite{2018ApJ...859L..25Y}, the authors studied the case of equal mass BBH. Here we inherent this simplification. The total mass is just twice the individual mass at coalescence to be calculated with equation (\ref{eqn:2018}): $M_{\rm{tot}}=2m_{\rm{f}}\,M_\odot$. 

The effective spin $\chi_{\rm{eff}}$ is defined as: 
\begin{equation}
    \chi_{\rm{eff}}=\frac{\left(M_1a_{*,1}\cos\theta_1+M_2a_{*,2}\cos\theta_2\right)}{M_{\rm{tot}}},
    \label{eqn:chi}
\end{equation}
where $M_i$ and $a_{*,i}$ ($i=1,2$) are the mass and spin parameter of individual BH respectively, $\theta_i$ is the angle between the spin and the orbital angular momenta. 

Given a distribution of initial mass, spin parameters, misalignment angles, Eddington accretion rate of individual BHs and the initial separation, we can calculate a distribution of the $\chi_{\rm{eff}}$ and $M_{\rm{tot}}$ at merger using equations (\ref{eqn:differential}-\ref{eqn:chi}). From a distribution of these parameters, the probability density of $\chi_{\rm{eff}}$ and $M_{\rm{tot}}$ at merger can be obtained using Monte-Carlo method. 

\begin{figure}
    \centering
    \includegraphics[width=0.45\textwidth]{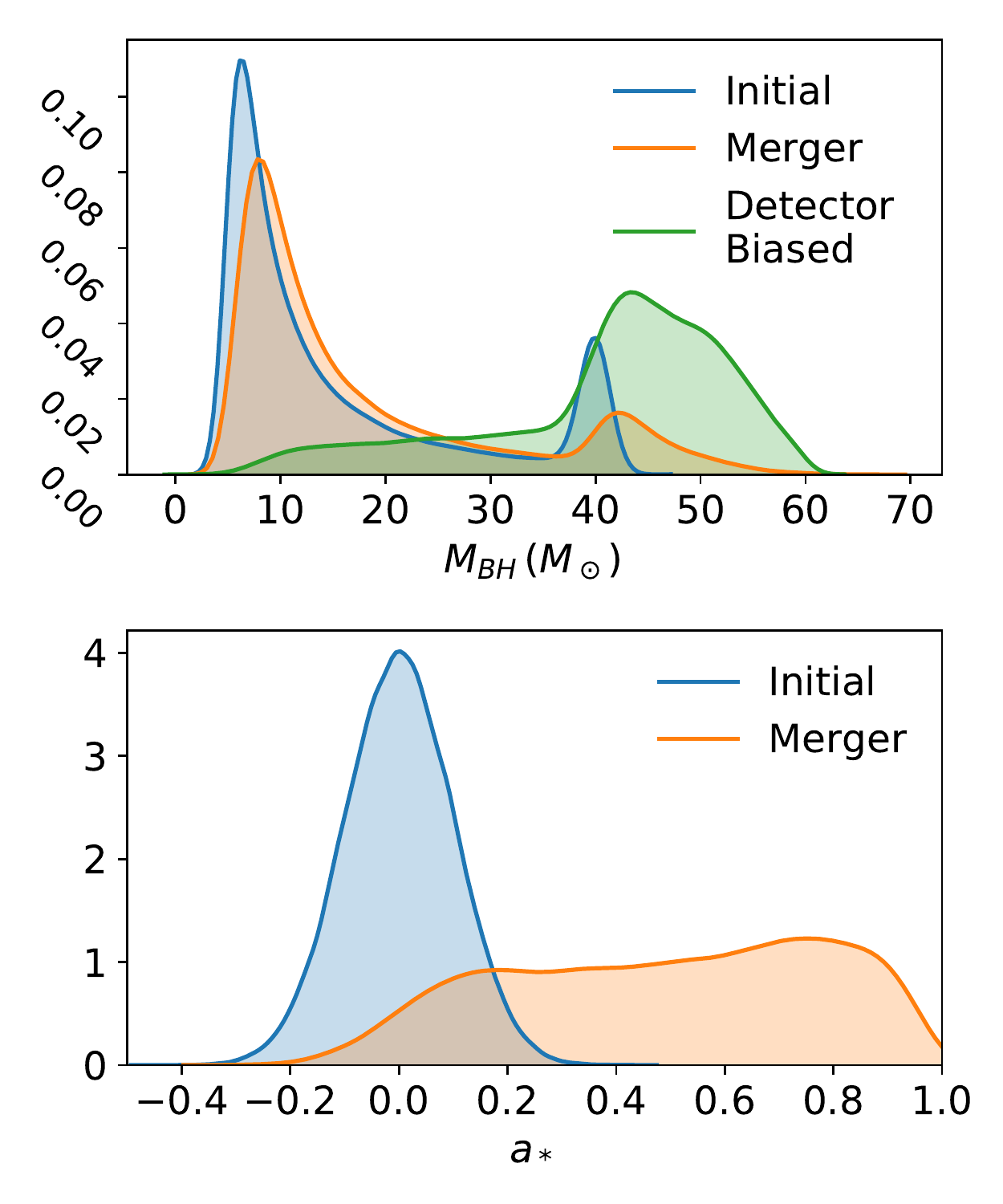}
    \caption{\textbf{Upper panel:} Probability distribution masses of individual BHs initially (blue), at merger (orange) and detector biased (green); \textbf{Lower panel:} Probability distribution of spin parameter initially (blue) and at merger (orange). }
    \label{fig:kde1d_1}
\end{figure}

\begin{figure}
     \centering
     \includegraphics[trim={3.3cm 0 2cm  0},clip,width=0.45\textwidth]{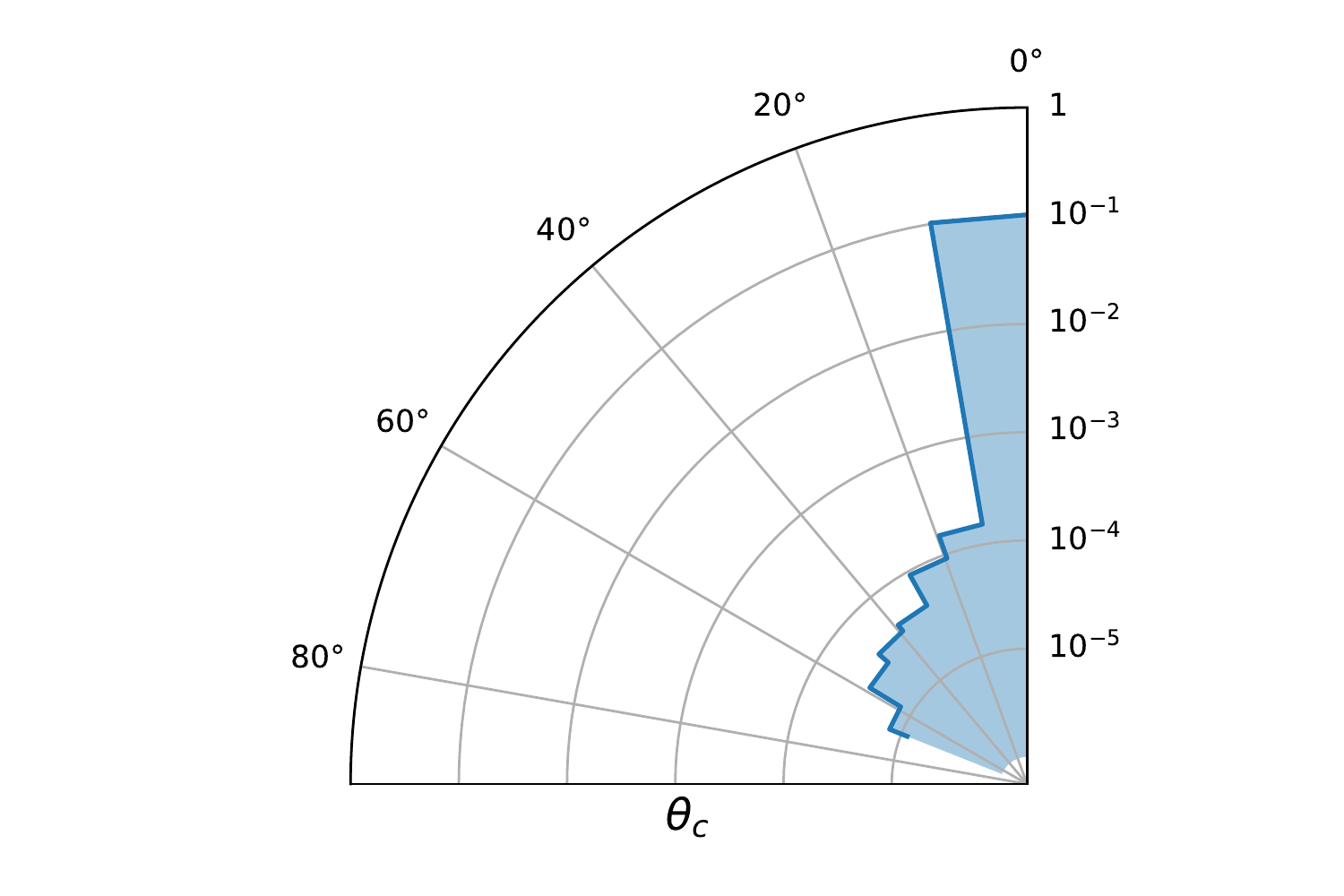}
     \caption{Normalized distribution of misalignment angle $\theta$ of BHs at merger. The radial axis is in logarithm scale. }
     \label{fig:polar}
 \end{figure}

We assume that the initial mass function of BHs in AGN disk channel is the same as that in clean channels. We use a parameterization proposed by \cite{2018ApJ...856..173T}: 
\begin{equation}
    p(M_0)=(1-\lambda)p_{\rm{pow}}(M_0)+\lambda p_{\rm{Gaussian}}(M_0),
\end{equation}
where $p_{\rm{pow}}$ is a power law with the index $-\alpha$, low mass cut-off at $M_{\rm{gap}}$ and high mass cut-off at $M_{\rm{cap}}$; $p_{\rm{Gaussian}}$ is a Gaussian peak. This peak distribution attributes BHs formed via pulsational pair-instability supernovae (PPSNe). We denote the mean and variance of the peak as $m_{pp}$ and $\sigma_{pp}$. $\lambda$ is the portion of PPSNe BHs, which can be estimated through:
\begin{equation}
    \lambda=\left(\frac{m_{\rm{gap}}}{m_{\rm{cap}}}\right)^{\alpha-1}.
\end{equation}
We take the fiducial parameters: $M_{\rm{gap}}=5\,M_\odot$, $M_{\rm{cap}}=40\,M_\odot$, $\alpha=2$, $m_{pp}=M_{\rm{cap}}$, $\sigma_{\rm{pp}}=1\,M_\odot$. The probability density function (PDF) of the initial mass of individual BH is plotted as blue line in the upper panel of figure \ref{fig:kde1d_1}. 
The distribution of other parameters as follow: $\gamma$ is calculated from $\tau_{\rm{sink}}$ as \citep{2017MNRAS.469.4258T}:
    \begin{equation}
    \gamma=-0.496+1.68\tau_{\rm{sink}},   
    \end{equation}
 and $\tau_{\rm{sink}}$ is sampled uniformly from 3 to 5 \citep{2018ApJ...859L..25Y}; $f_{\rm{Edd}}$ is sampled uniformly in log-space from -3 to 1; For the distribution of the binary separation $a_{12,0}$, we use the same in \citep{2018ApJ...859L..25Y}, i.e., \"Opik's law with an upper limit from ionization of the binary, with the upper limit of $a_{12,0}=5$. 
The PDF of individual BH masses at merger is plotted as orange line in the upper panel of figure \ref{fig:kde1d_1}. The initial $a_*$ of individual BHs is assumed to follow a Gaussian distribution centering at zero and with a width 0.1, which is plotted together with the pdf of $a_*$ at merger in the lower panel of figure \ref{fig:kde1d_1}. The initial misalignment angle is assumed to be isotropic. Nearly all BHs have their spin axis align with the orbital axis at merger, as shown in figure \ref{fig:polar}. The time of growth is calculated with:
\begin{equation}
    \tau_{\rm{growth}}=\log(x)/f_{\rm{Edd}}\times10^8\,\text{yrs},
\end{equation}
We exclude those samples whose $\tau_{\rm{growth}}$ are larger than the typical lifetime of
AGNs, $\tau_{\rm{AGN}}$. Here we use a fiducial value of $\tau_{\rm{AGN}}=10^8$\,yrs.

We also want to include the effect that the detectable volume increasing with the chirp mass of BBH: $V\propto\mathcal{M}^{5/2}$ (in the local Universe, where $\mathcal{M}$ is the chirp mass of the BBH and it is proportional to $M_{\rm{tot}}$). As a result, the detected distribution will bias towards higher total mass:
 \begin{equation}
     p_{\rm{det}}(M_{\rm{tot}},\chi_{\rm{eff}})\propto p(M_{\rm{tot}},\chi_{\rm{eff}})M_{\rm{tot}}^{5/2}.
 \end{equation}
The detector biased mass PDF of BHs is plotted as green line in the upper panel of figure \ref{fig:kde1d_1}. 

\begin{figure}
    \centering
    \includegraphics[width=0.48\textwidth]{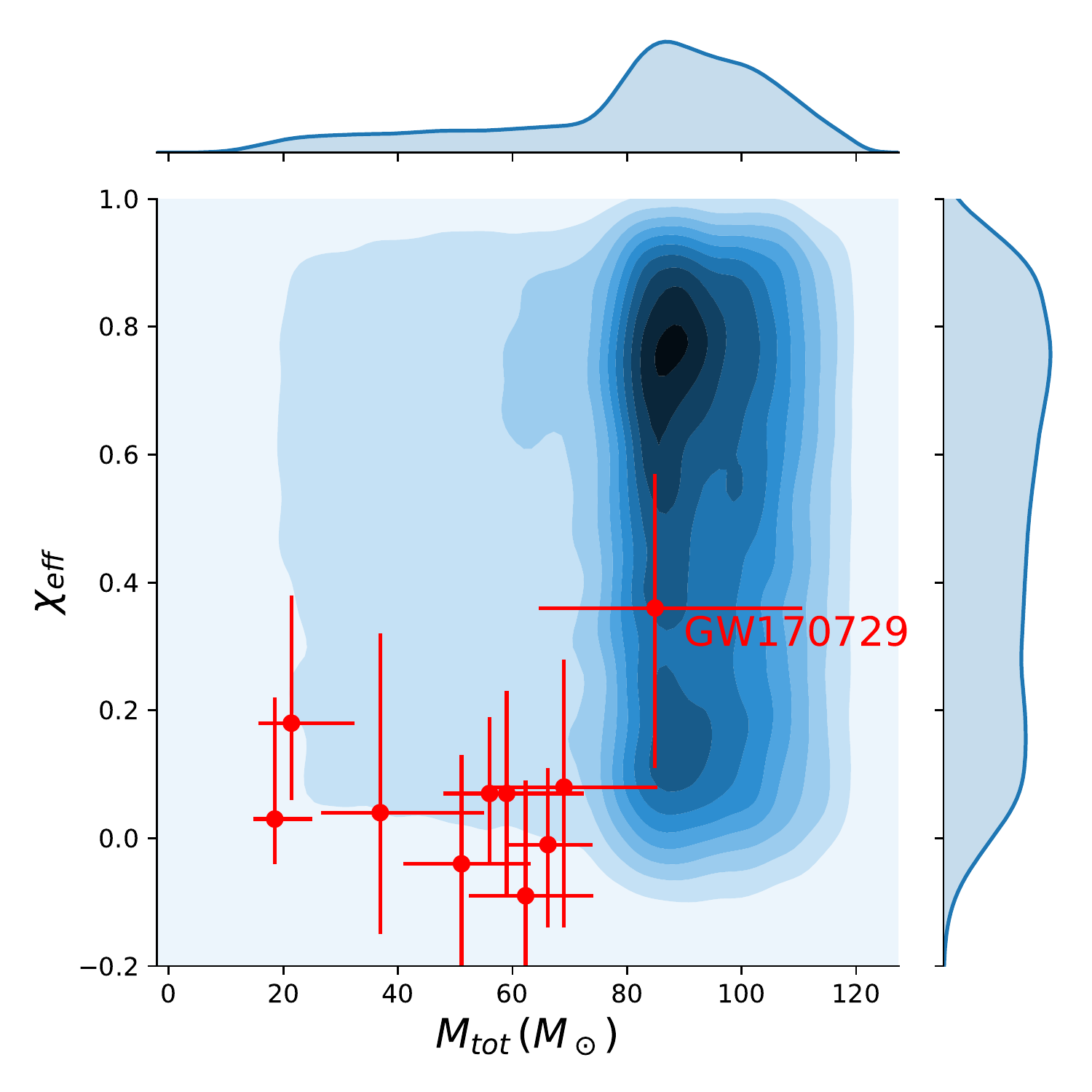}
    \caption{Simulated probability density distribution of the BBH from AGN-disk. The red points with error bars are data from LIGO/Virgo O1-O2 catalog.}
    \label{fig:2Dkde}
\end{figure}

In figure \ref{fig:2Dkde}, the density plot represents the probability density of detecting a AGN-disk channel BH with $M_{\rm{tot}}$ and $\chi_{\rm{eff}}$ with GW observation. We also plot the observed data with their uncertainty in the same figure.

It is shown that GW170729 falls far from the cluster of other events in the parameters space, and towards the peak of the predicted distribution of the AGN-disk channel events. It hints that GW170729 is from different channel with others. \cite{2017MNRAS.464..946S} found that BBH merger rate density in the AGN-disk channel is $\sim3$\,yr$^{-1}$\,Gpc$^{-1}$ with large uncertainty. It implies among all BBH merger GW events, the ratio of AGN-disk events can be a few percents to tens of percents (the BBH merger rate calcualted from LIGO/Virgo O1-O2 runs is $9.7-101$\,yr$^{-1}$\,Gpc$^{-1}$ according to \citealt{2018arXiv181112907T}). We suggest a search in the archival data for a FRB around the time of GW170729, and we expect to see some more AGN-disk channel candidates after the O3 run. 

\section{Discussion}\label{sec:discussion}
\subsection{Observation of EM counterparts and host galaxies}
It is possible that a FRB will be observed accompanying a GW chirp event, if the BBH merger is viewed from face-on. With more GW observatories joining the network, the inclination of the BBH will be constrained with increasing accuracy. Due to the extreme short duration of the FRB, the counterpart of GW is only expected to be detected by chance when the field of view (FoV) of a radio telescope covers the right GW sources, i.e., BBH mergers from the AGN-channel, with small orbital inclination angle. Since the FoV of FRB monitors are usually large (e.g., Canadian Hydrogen Intensity Mapping Experiment (CHIME) has the FoV of $\sim200$\,deg$^2$ \citealt{2018ApJ...863...48C}), it is still likely to have simultaneous detection of GW and the FRB. 

The candidates of the host galaxy will be chosen from a sample of type II AGNs, from a sky region jointly constrained by the FRB and GW observation, and within a certain red shift range. 
\subsection{Caveats}
In section \ref{sec:parameterspace}, we demonstrated that comparing with BBH merger from clean channels, ones from the AGN-channel have higher possibility towards large $M_{\rm{tot}}$ and $\chi_{\rm{eff}}$. Changing the distribution of input parameters will not qualitatively change this conclusion. However, our result cannot be used quantitatively to predict the possibility that a certain GW event is from AGN-disk channel other than clean channels. It is because that there are large uncertainties in the distribution of parameters determining the $M_{\rm{tot}}-\chi_{\rm{eff}}$ probability density. The relative ratio among different BBH formation channels are also not well constrained. 

We assumed that the initial mass function and spin distribution of BHs in the AGN channel is the same with those in clean channels. The reality could be different from our assumptions: \cite{2019ApJ...876..122Y} showed that the initial mass function was  hardened by the AGN disk due to the orbital alignment process. A non-negligible fraction of BHs experienced previous mergers (hierarchical mergers), which resulted in heavier mass and faster aligned spin \citep{2019arXiv190609281Y,2019arXiv190704356M}. We hope to include the above-mentioned factors in future studies.
\acknowledgments
SXY acknowledge support from the Netherlands Organisation for Scientific Research (NWO). KSC is supported by a GRF grant under 17310916. 


\end{document}